\documentclass[twocolumn,graphicx,reprint]{revtex4-1}
\usepackage{float}
\usepackage{graphicx}
\usepackage{gensymb}
\usepackage{siunitx}
\usepackage{upgreek}
\usepackage{natbib}
\begin{document}
\raggedbottom

\title{Efficient Photon Coupling from a Diamond Nitrogen \\ Vacancy Centre by Integration with Silica Fibre}

\author{Rishi N. Patel$^{1,2}$}
\email{rishipat@stanford.edu}

\author{Tim Schr{\"o}der$^{1}$}
\email{schroder@mit.edu}

\author{Noel Wan$^{1}$}

\author{Luozhou Li$^{1}$}

\author{Sara L. Mouradian$^{1}$}

\author{Edward H. Chen$^{1}$}
\author{Dirk R. Englund$^{1}$}
\email{englund@mit.edu}
\affiliation{$^{1}$Department of Electrical Engineering and Computer Science, MIT, 77 Massachusetts Ave. Cambridge MA, 02139}
\affiliation{$^{2}$Department of Applied Physics, Stanford University, 348 Via Pueblo Mall, Stanford CA, 94305 }

\date{\today}

\begin{abstract}

A central goal in quantum information science is to efficiently interface photons with single optical modes for quantum networking and distributed quantum computing. Here, we introduce and experimentally demonstrate a compact and efficient method for the low-loss coupling of a solid-state qubit, the nitrogen vacancy (NV) centre in diamond, with a single-mode optical fibre. In this approach, single-mode tapered diamond waveguides containing exactly one high quality NV memory are selected and integrated on tapered silica fibres. Numerical optimization of an adiabatic coupler indicates that near-unity-efficiency photon transfer is possible between the two modes. Experimentally, we find an overall collection efficiency between 18-40\% and observe a raw single photon count rate above 700 kHz. This integrated system enables robust, alignment-free, and efficient interfacing of single-mode optical fibres with single photon emitters and quantum memories in solids.
\end{abstract}


\maketitle %\maketitle must follow title, authors, abstract and \pacs

%%%%%%%%Introduction

%%%%%%%%%
%\section{Introduction}
Efficient coupling of stationary quantum memories to a single spatial mode in silica fibre is of central importance in a range of quantum information processing applications, including long-distance entanglement of stationary quits and quantum networks \cite{Kimble2008,Northup2014_Nature,Childress2013_MRS,Loncar2013_MRS,sprague2014broadband}. Recently, the efficient fibre coupling of atomic quantum memories also enabled large atom-cavity coupling \cite{thompson2013coupling,tiecke2014nanophotonic,TieckeFiber} and strong single-atom nonlinearities \cite{Shea2013fiber,mitsch2014directional}. Among solid state qubits, the nitrogen-vacancy (NV) centre in diamond has emerged as an attractive quantum memory due to its optically addressable and long-coherence electronic and nuclear spin states \cite{Maurer2012_science,Bar-Gill2013_electronSpinCoherence}. These properties have enabled recent demonstrations of heralded quantum entanglement \cite{Bernien2012_Nature_heraldedEntanglement} and teleportation \cite{pfaff2014unconditional} between two separated NV centres. To improve the entanglement probability in such schemes, an open experimental challenge is to improve the efficiency with which single photons from an NV centre can be channeled into a single guided optical mode. This has motivated a variety of light collection approaches, including diamond micro posts \cite{Babinec2010}, solid-immersion-lenses \cite{schroder2011ultrabright,Hadden_etal,Siyushev_etal}, grating structures \cite{choy2013spontaneous,li2014three}, and inverse tapered coupling to photonic integrated circuits \cite{Hausmann2012_nanoletters_NVringResonator}. Several research efforts have also sought to integrate quantum emitters directly with optical fibres, as a way of eliminating non-essential optical components and achieving compact and nearly monolithic interfaces. Recently, fluorescence collection from colloidal quantum dots and diamond nanocrystals containing single NVs was proposed \cite{Chonan2014_natureSciReports,almokhtar2014numerical} and demonstrated, often using a tapered fibre section \cite{Yalla2012_PRL_QDonTaper,Schroder2011_NanoLett_PCfiber,Schroder2012_NDonTaper_OptExp,Liebermeister2014,Liu2013_APL,fujiwara2011highly}. However, these approaches use point-like emitters that exhibit poor collection efficiency; in addition, the spin and optical properties of NVs in diamond nanocrystals are degraded compared to bulk diamond \cite{almokhtar2014numerical}. Here, we introduce an approach that solves these problems by adiabatic power transfer between a tapered silica fibre and a single-mode diamond micro-waveguide fabricated from high-quality CVD-grown diamond. This integrated diamond-silica waveguide system enables efficient optical collection from high-quality NV centres.   
	
\begin{figure*}[t!]
 \centerline{\includegraphics[scale=.5]{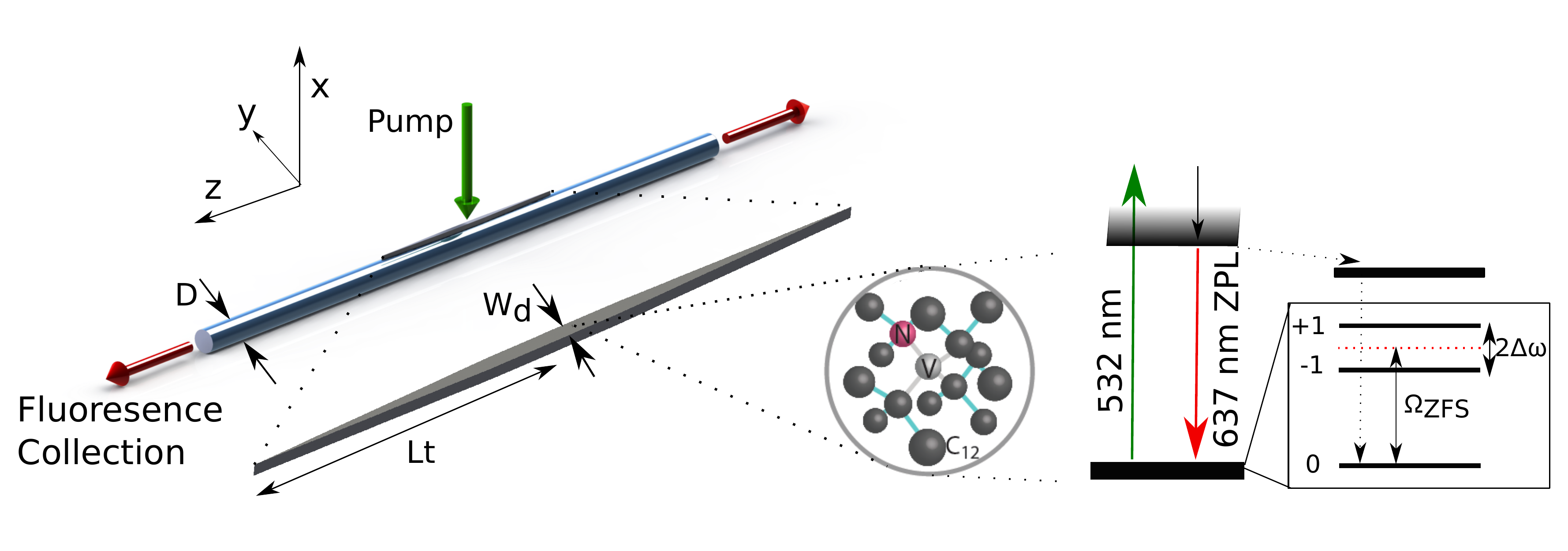}}%
 \caption{\textbf{System Overview} Schematic showing a diamond micro-waveguide containing an NV centre (grey) in parallel contact with a tapered fibre segment (blue). Excitation and collection pathways are shown in green and red respectively. The inset shows the diamond crystal lattice and the NV centre spectrum. The parameters in this experiment are the fibre diameter $D \approx 500$ nm, the diamond width (at z = 0), $W_{d}(0) = 200$ nm, and the taper length $L_{t} = 5$ $\upmu$m. In the NV energy level diagram, the zero phonon line is denoted by ZPL, the zero-field-splitting frequency of 2.87 GHz by $\Omega_{ZFS}$  and the Zeeman splitting between the +1 and -1 states in a DC magnetic field by $2\Delta\omega$.}%
 \label{fig:overview}
 \end{figure*}

%\section{Results}
%\subsection{Device Description}
We consider a tapered single-mode diamond micro-waveguide in direct contact with the tapered section of a single-mode optical fibre, as illustrated in Figure \ref{fig:overview}. The diamond micro-waveguide is positioned by van der Waals forces directly on the tapered fibre resulting in coupling between the optical modes of the two structures.

%Fig1
%%%%%%%%%%%%%%%%%%%%
 %%%%%%%%%%%%%%%%%%%

\begin{figure*}[t!]
 \centerline{\includegraphics[scale=.9]{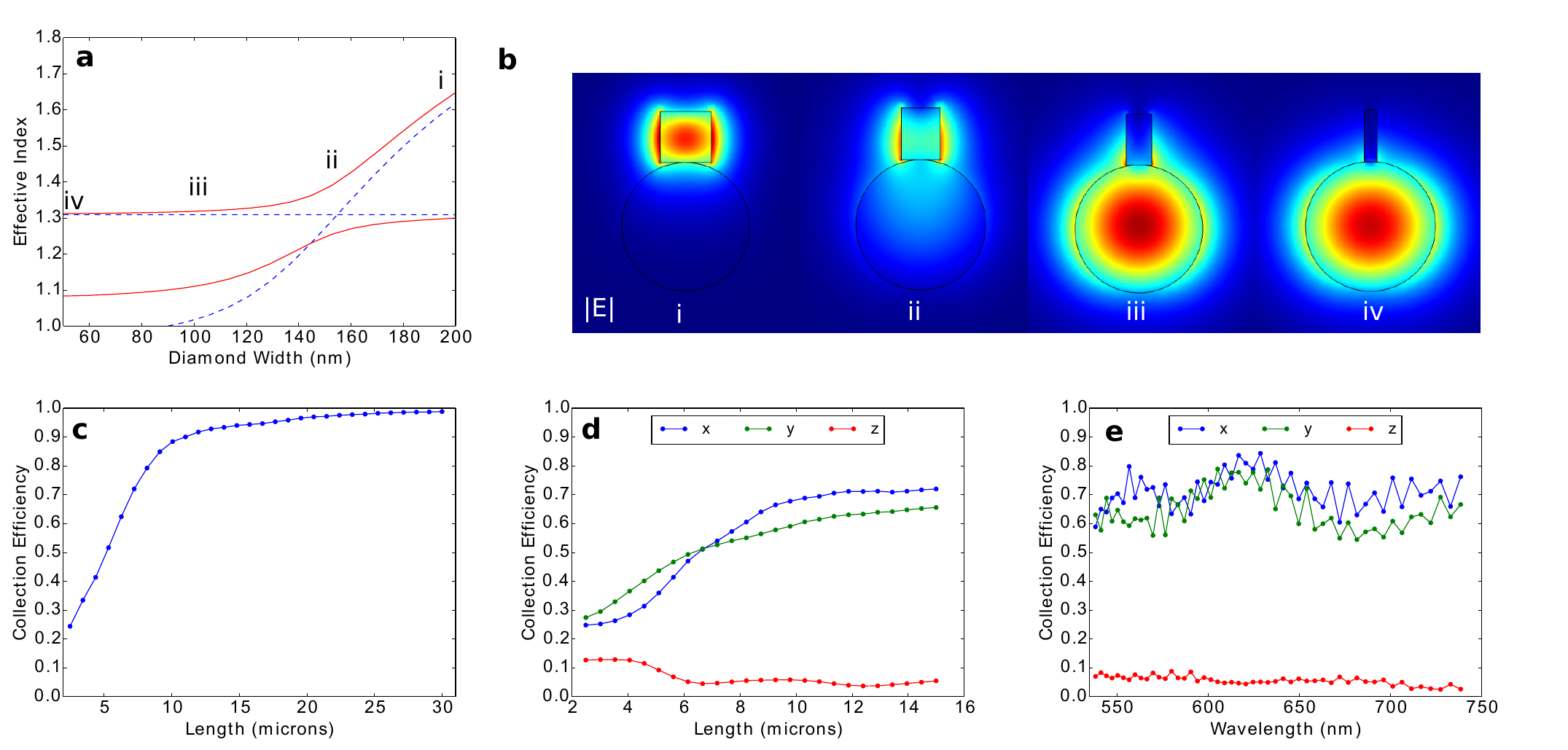}}%
 \caption{\textbf{System Simulation} \textbf{(a)} Plot of the effective mode index of the fundamental mode of the fibre and diamond micro-waveguide versus diamond width. The blue dotted lines show the uncoupled fundamental modes of each structure. The horizontal line is the effective index of the lowest order fibre mode. The red lines are the coupled supermodes, which show an anti crossing. \textbf{(b)} The field profile of the norm of the electric field is shown for the fundamental supermode of the combined structure. From left to right, the diamond width ranges from 200 nm, 150 nm, 100 nm, and 50 nm. \textbf{(c)} FDTD calculation using a mode source to excite the fundamental supermode. Collection efficiency approaches unity (over 98 \%) \textbf{(d)} FDTD calculation using a dipole source at three different polarizations. Blue, green and red are oriented along the x, y and z directions, respectively. See Figure 1 for a coordinate inset. \textbf{(e)} Detailed view of the FDTD results vs. wavelength for a 15um long diamond micro-waveguide. The polarization colour coding is the same as in \textbf{(d)}.}%
 \label{fig:simulations}
 \end{figure*}

For a slowly varying diamond width, corresponding to a slowly varying effective refractive index, light in the diamond micro-waveguide remains in the fundamental mode of the combined diamond-silica structure. We analyse this problem using coupled mode analysis, following a similar analysis for coupling between a silicon waveguide to a tapered fibre\cite{Groblacher2013_adiabaticCoupling_APL}. Figure \ref{fig:simulations}a plots the effective indices of the diamond and fibre modes as a function of the diamond waveguide cross section, for a fixed fibre diameter of 500 nm. Near a diamond cross section of 140 nm, a clear anti-crossing is observed as the matching group indices of the waveguides results in strong mode-coupling between them. The corresponding mode fields are shown in Figure 2b. If the diamond taper is swept slowly such that the optical mode remains in the ground state through the transition from the diamond into the silica waveguide, nearly 100 \% power transfer is possible. Specifically, the diamond taper width $W_d(z)$ must satisfy: 
\begin{equation}
\frac{d W_{d}(z)}{dz} \ll  \left(\frac{d n_{d}}{d W_{d}(z)}\right)^{-1} \frac{2 \pi}{\lambda_{o}} (\Delta n_{eff})^{2}
\end{equation}

where $\frac{d n_{d}}{d W_{d}(z)}$, the rate of change of the fundamental diamond mode effective index versus diamond width, can be computed from the derivative of the uncoupled diamond band plot shown in Figure 2a. The parameters on the right-hand-side of Eqn. (1) can be estimated from the simulation shown in Figure \ref{fig:simulations}a. Using $\lambda_{o} = 637$ nm, $\Delta n_{eff} = 0.13$ and $\frac{d n_{d}}{d W_{d}(z)} =  0.008$, we thus obtained a bound $\frac{d W_{d}(z)}{dz} < 0.02$. For $W_d(0)$ = 200 nm this corresponds to a taper length $L_{t} = 10 $ $\upmu$m. This is consistent with the FDTD calculation shown in Fig. \ref{fig:simulations}c, confirming that the coupling efficiency increases with the length of the diamond waveguide, as expected. 

	Next, we consider how the orientation of the NV, which is assumed to be located at the centre of the diamond micro-waveguide (see Figure 1), affects the coupling efficiency. In our simulation we consider three orthogonal dipole polarizations along the axes shown in the inset of Figure 1a. For an x-polarized dipole, a theoretical overall collection efficiency of over 70\% can be achieved, as can be seen from the collection efficiency plot for the three dipole orientations in Figure \ref{fig:simulations}d. We also note that the adiabatic coupling approach has the advantage of being highly broadband, as is apparent from the FDTD calculations plotted versus excitation wavelength (Fig. \ref{fig:simulations}e).

 %Fig2
 %%%%%%%%%%%%%%%%%%%%
 %%%%%%%%%%%%%%%%%%%%

\begin{figure*}[t!]
 \centerline{\includegraphics[scale = 0.9]{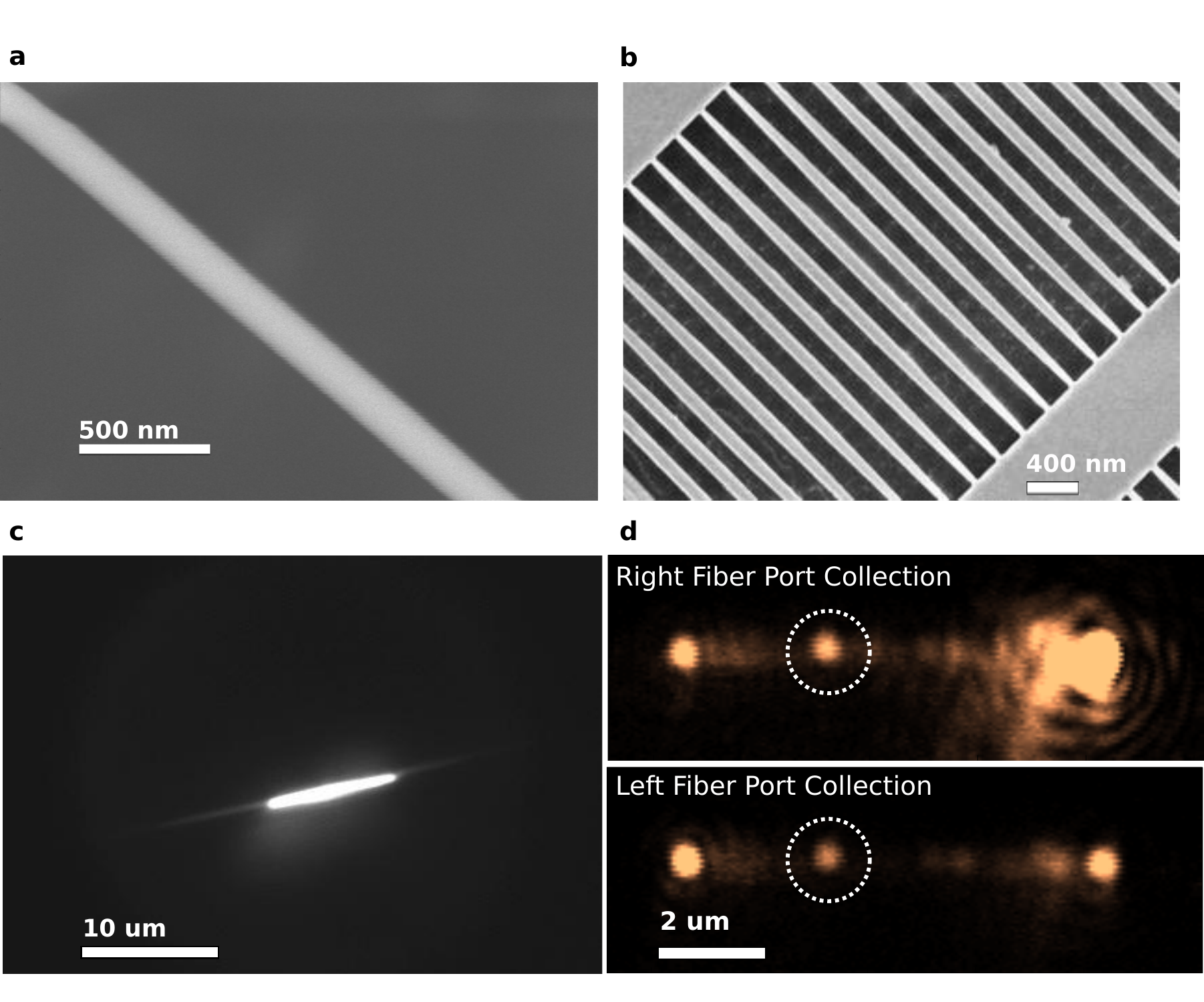}}%
 \caption{\textbf{Device Fabrication} \textbf{(a)} Tapered fibres with sub-wavelength diameter waists are fabricated using a heat and pull method \cite{fiberTapering}. \textbf{(b)} Scanning electron microscope image of diamond micro-waveguides. \textbf{(c)} Optical image of assembled device, consisting of a micro-waveguide placed onto a tapered fibre waist. \textbf{(d)} Confocal scans of fluorescence detected from two fibre ends (shown in top and bottom panel). The micro-waveguide is optically pumped with a green (532 nm) laser excitation source. Light originating from a single NV centre is shown in the dotted circles.}%
 \label{fig:experimental}
 \end{figure*}

The system fabrication involves three main stages. First, tapered fibres were fabricated from single-mode optical fibres (Thorlabs 630HP) using a standard heat-and-pull technique \cite{fiberTapering} while secured in a tapered fibre mount (Fig. \ref{fig:experimental}a) that is sealed to maintain a clean-air environment and to block air flow. The fibre tapering results in an adiabatic fibre mode conversion to maintain a high ($\approx  90 \%$) transmission of 633 nm laser light. Although especially long tapers can provide efficiencies around 97\% \cite{Vetsch2010_laserTrappedNanofiber_PRL}, we used here shorter tapers as they are easier to manipulate. Second, diamond micro-waveguides were fabricated from a 200 nm thin film of [1 0 0] electronic grade synthetic diamond using transferred hard mask lithography \cite{liSciReportsDiamond2015}. They are 12 $\upmu$m long, with a 5 $\upmu$m triangular taper on either side of a 2 $\upmu$m section of constant width (Fig. \ref{fig:experimental}b). Third, we characterized and selected individual diamond micro-waveguides containing exactly one NV centre before placing them onto the waist of the tapered fibre using a nano-manipulator system, where they adhere readily by van der Waals forces. Fig. \ref{fig:experimental}c shows a completed device. In a confocal scan, the NV can be identified as a bright fluorescence spot to the left of the centre of the  micro-waveguide (Fig. \ref{fig:experimental}d). Here, we excited the NV from the top while collecting photons through the left and right fibre ends, as seen in the top and bottom panels, respectively. The ends of the diamond micro-waveguides are also apparent as bright fluorescence spots on the scan. At these positions, more laser power is scattered into the fibre, causing increased fluorescence background. The bright extended spot at the right of the top panel is due to additional asymmetric scattering from the tip of the diamond waveguide. 

%%%%%%%%%%%%%%%%%%%%
 %%%%%%%%%%%%%%%%%%%%
%\sbsection{Optical Characterization}

\begin{figure*}[t!]
 \centerline{\includegraphics[scale = 0.94]{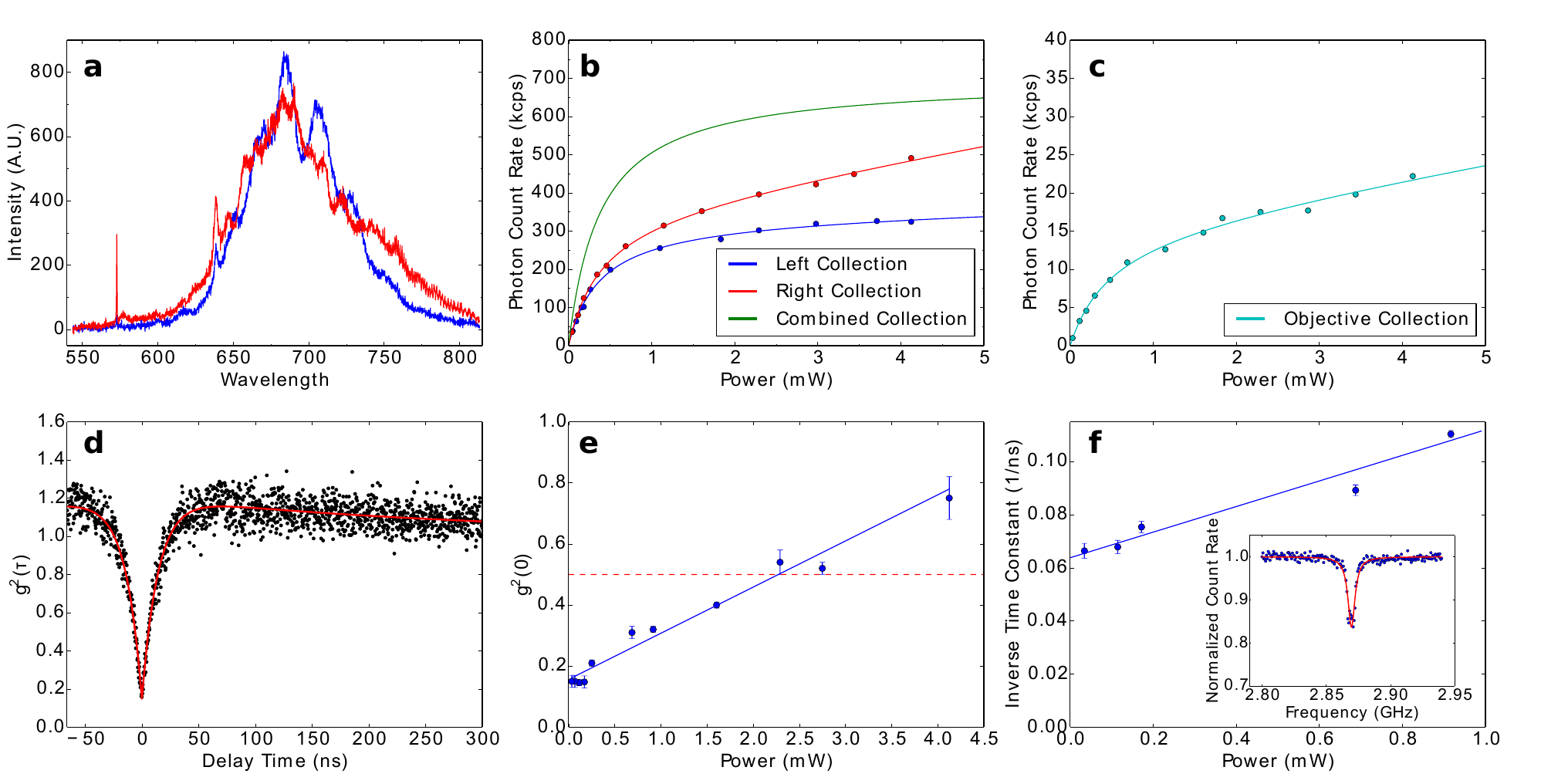}}%
 \caption{\textbf{Optical Characterization of Tapered-fibre Coupled Diamond Micro-waveguides.} \textbf{(a)} Spectra in blue (red) show NV fluorescence collected through left (right) ends. The zero phonon line and sideband are clearly detected. We observe stronger signal from the right side, as both spectra are taken under the same integration time. \textbf{(b)} Saturation curves showing photon count rates from NV centre versus pump power collected through the fibre. The blue (red) shows collection from the left (right). \textbf{(c)} Saturation curve data using objective collection with an NA  = 0.75 objective.  The green curve is the combined signal from both ends (with linear background subtracted), and hence the total number of guided single photons. \textbf{(d)} Second order autocorrelation function taken using both fibre ends in a Hanbury-Brown-Twiss setup. Strong photon anti-bunching with $g^{(2)}(0) < 0.2$ is observed.  \textbf{(e)} Strong photon anti-bunching is detected with $g^{(2)}(0) < 0.5$ over a range of powers below about 2.3 mW. The blue line is a guide to the eye, while the red dotted line denotes 0.5. \textbf{(f)} Measurements of the $g^{(2)}(\tau)$ time constant (describing the central dip at low pump power) fitted to a linear model yield a lifetime of 15.7 ns. (Inset) Electron-spin-resonance showing a fluorescence dip at the zero field splitting frequency of 2.87 GHz}%
 \label{fig:opticalMeasurements}
 \end{figure*}

Optical measurements reveal the efficient collection of NV fluorescence directly coupled into a single guided mode. We detected NV fluorescence through both the left and right fibre ends, as evidenced by the typical NV spectrum with a pronounced zero phonon line around 637 nm (Fig. \ref{fig:opticalMeasurements}a). 

%Fig.4	
%%%%%%%%%%%%%%%%%%%%	
 %%%%%%%%%%%%%%%%%%%%
 
	By varying the pump power, we observe a saturation behavior in the NV signal, (Fig. \ref{fig:opticalMeasurements}b,c). All fits are to a model of the form:
\begin{equation}
I(P) = \frac{P I_{sat}} {(P_{sat} + P)} + C P 
\end{equation}
where $I$ is the measured intensity in counts per second (cps), $P$ is the pump power, $I_{sat}$ the saturation intensity, $P_{sat}$ the saturation pump power, and $C$ is a fitting parameter that characterizes the linear background contribution. The fits estimate an overall $I_{sat}$ of $(712 \pm 24) \cdot 10^{3}$ cps  from both fibre ends. We measured the normalized second order autocorrelation function of the fibre-coupled light with a Hanbury-Brown-Twiss setup, using the fibre itself as the intrinsic beamsplitter. The anti-bunching with $g^{(2)}(0) = (0.15 \pm 0.02) < 0.5$ indicates the presence of a single NV centre; no background was subtracted from this measurement (Fig. \ref{fig:opticalMeasurements}d). This autocorrelation data is fit to a three-level model,
\begin{equation} \label{eq:g2norm}
g^{(2)}(\tau) = 1 + p^{2}_{f} ( c e^{-|\tau|/\tau_{1}}  - (1+c) e^{-|\tau|/\tau_{2}}),
\end{equation}
where $p_{f}$ is the single photon emission probability, $\tau_{1}$ is a time constant for the central anti-bunching dip, $\tau_{2}$ is the time constant for correlation function decay at intermediate times, and $c$ is a parameter describing the amplitude of the bunching shoulders. We then estimated the background by plotting the parameter $g^{(2)}(0) = 1 - p^{2}_{f}$ for different values of the excitation pump power (Fig. \ref{fig:opticalMeasurements}e). Here we identify a maximum incident pump power of 2.25 mW focused to a spot size of approximately 0.6 $\upmu$m, beyond which the background emission causes $g^{(2)}(0) > 0.5$. At this pump power, we use the green curve in Fig. \ref{fig:opticalMeasurements}b to show that over 600,000 single photons per second are detected. Fitting Eqn. (3) for a number of different pump powers allows us to extract the inverse time constant, $1/\tau_{1}$ (Fig. \ref{fig:opticalMeasurements}f). We see that $1/\tau_{1}$ varies linearly for small pump power, and extrapolation to zero power yields a lifetime, $\tau_{NV}$, of $15.7 \pm 1.1$ns, in good agreement with previous studies on NVs in diamond nanowires \cite{Babinec2010}.

%\subsection{System Efficiency}

We use two different metrics to characterize our observed system efficiency. The first, relating the number of guided photons to the total possible number of photons emitted with a unity quantum efficiency, is given by,
\begin{equation} \label{eq:betaDef}
\beta = \frac{\Gamma_{guided}}{\Gamma_{decay}},
\end{equation}
where $\Gamma_{guided}$ is the total excited state decay rate into the guided modes (into the fibre), and $\Gamma_{decay}$ is the total decay rate of the NV centre (including both non-radiative and radiative pathways). $\beta$ is a lower bound on the observed system efficiency if one assumes an NV quantum efficiency of unity. We estimate the parameter $\Gamma_{guided}$ from our measurements of the observed count rate,
\begin{equation} \label{eq:gammaGuided}
\Gamma^{*}_{guided} = \eta_{apd} \,  \kappa_{tg} \, \kappa_{taper} \, \Gamma_{guided},
\end{equation}

where $\eta_{apd}$ is the quantum efficiency of our avalanche-photo-diode (APD) detectors, $\kappa_{taper}$ the transmission efficiency of our tapered fibre, and $\kappa_{tg}$ the transmission efficiency of our filtering stages. In our experiment, $\eta_{apd} \approx 65\%$, $\kappa_{taper} \approx 95\%$ and $\kappa_{tg}  \approx 43\% $. $\kappa_{taper}$ is obtained from transmission measurements of the fibre during fabrication, and represents an upper bound on the taper transmission. It is defined as the single-sided transmission, from the centre of the taper to either of the fibre ends. $\kappa_{tg}$ is obtained from measuring the transmission of a 635 nm laser signal through both left and right filtering ports, providing an upper bound for the filter stage transmission. We take a weighted average of these two measurements for our estimated $\kappa_{tg}$ and $\Gamma^{*}_{guided} = (712 \pm 24) \cdot 10^{3}$ cps directly from the parameter $I_{sat}$ above. Dividing out the transmission factors gives a total single photon count rate of about $\Gamma_{guided} = (2.7 \pm 0.1) \cdot 10^{6}$ cps . We determine $\Gamma_{decay}$ as the inverse of the lifetime, $15.7 \pm 1.1$ ns.  Finally, this gives a value of $\beta \approx 4.2\%$. This lower bound of $\beta$ does not take into account the NV's charge instability or the non-unity quantum efficiency. 

In a second method of estimating the collection efficiency, we use the ratio of the guided photons to the total number of radiated photons,
\begin{equation} \label{eq:colleffDef}
\eta_{c} = \frac{1}{1 + \frac{\Gamma_{free}}{\Gamma_{guided}}},
\end{equation}
where $\Gamma_{free}$ is the decay rate into free space modes. $\eta_{c}$ differs from the first metric, because it requires an estimation of the total number of radiated photons as opposed to the total decay rate, which includes both non-radiative and radiative terms. That is, we can write $\eta_{c} = \beta / \eta_{QE}$, where $\eta_{QE}$ is the quantum efficiency of the NV. 
We estimate the free space decay rate from
\begin{equation} \label{eq:gammaFree}
\Gamma^{*}_{free} = \eta_{apd}  \, \kappa_{tf} \, \kappa_{NA} \,  \Gamma_{free}
\end{equation}
where $\kappa_{tf}$ is the transmission efficiency of our confocal microscope setup and $\kappa_{NA}$ gives the fraction of photons collected into the acceptance angle of our microscope objective (NA = 0.75, cover-glass corrected Zeiss). This fraction is not easily obtained since the dipole orientations of our emitter are not known \textit{a priori}. To estimate the NV orientation in our diamond micro-waveguide, we performed polarization-dependent excitation measurements and fit the resulting fluorescence data assuming two orthogonal NV radiation dipoles (see Supplementary Information)\cite{dolan2014complete_polarizationRadiallyPolarizedBeams_NV}\cite{PhysRevB.76.165205_PolarizationSelectiveExcitation_AlegreSantori}. With FDTD simulations, we then obtained $\kappa_{NA}$. We also estimated a range for $\kappa_{tf}$ based on transmission measurements of our confocal setup, and taking into account losses from the cover glass enclosing our sample. From saturation measurements acquired via the objective, we have $\Gamma^{*}_{free} = (16 \pm 2) \cdot 10^{3}$ s $^{-1}$. Using the factors $\kappa^{low}_{tf} \approx 1\%$, $\kappa^{high}_{tf} \approx 3\%$, $\kappa_{NA} \approx 20\%$, we estimate an efficiency of  $\eta^{lower}_{c} \approx 18.1\%$ and $\eta^{upper}_{c} \approx 39.8\%$. The large range of $\kappa_{tf}$ is caused by our collection optics, as the fibre coupling efficiency is strongly dependent on the position of the sealing cover slip in the focus, and the setting of the correction ring of the objective. By comparison, our FDTD-based calculation yields a lower and upper bound on the collection efficiency of 0.49 and 0.52 respectively. We attribute the discrepancy primarily to the assumption in our simulation of a triangularly shaped diamond micro-waveguide tapered with ends going to zero width, an assumption that is not valid in the experimental system. In addition, the NV is not located exactly in the centre of the adiabatic coupler, and may have different coupling efficiencies to either side. In practice, we can achieve micro-waveguide tips of about 50 nm which creates a sudden index step at which scattering losses can occur. Using the above estimates for the collection efficiency, we obtain quantum efficiency bounds of $\eta^{lower}_{NV} \approx 10.6 \%$ and $\eta^{upper}_{NV} \approx 23.3 \%$. These values are in agreement with recent measurements in nanodiamonds \cite{Liebermeister2014}. 

The inset of Figure \ref{fig:opticalMeasurements}f shows continuous wave optically detected magnetic resonance (ODMR) of our NV electron spin. Separate Hahn-echo measurements on waveguides produced in the same way indicate long phase coherence times in excess of 100 $\upmu$s (see SI)\cite{saraNWwaveguide}. 
%\section{Discussion}

	It is important to cover several limitations of our system, and to highlight some aspects for future work. We observed that the tapered fibre transmission degrades over time because of the deposition of large scatterers, such as dust. In future experiments this can be averted by mounting the tapered section in a sealed container. Furthermore, presently our method is limited by diamond fabrication capabilities to make slowly tapering structures, as we currently attach the waveguides on the ends to a diamond substrate. With improved fabrication techniques, diamond waveguides could be made with longer taper lengths and thinner tips, improving adiabatic power transfer up to the theoretical limit close to unity. Regarding our optical pumping scheme, we currently use confocal excitation to excite individual NV centres in our devices. For a completely fibre-integrated approach, it would be interesting to use a fibre-integrated Bragg filter to allow excitation from one fibre end and collection through the second fibre end. Such a system could be cooled by simple immersion in a cryogenic fluid, eliminating the need for a confocal microscope and in principle allowing for lifetime-limited ZPL emission, an important requirement for a number of quantum information experiments. 

	In conclusion, we have demonstrated a novel approach for the efficient integration of a high-quality quantum memory directly with a silica fibre. With a raw count rate exceeding $7 \cdot 10^{5}$ cps and $g^{(2)}(0) \ll 0.5$,  we show a roughly four fold improvement over previous fibre-coupled approaches that used diamond point emitters instead of wave guiding structures \cite{schroder2011ultrabright}. In addition, the fibre background is strongly suppressed allowing exceptionally low $g^{(2)}(0) \approx 0.15 $ (without background subtraction) for cross-correlation measurements in a fibre-integrated system. This geometry can be used for coherent spin manipulation of a fibre-coupled spin qubit, thereby providing efficient optical access to a long-lived quantum memory. The deterministic pick and place method presented here is amenable to larger scale integration and can be extended to the evanescent integration of diamond-based nano-cavities \cite{li2014coherent}. Furthermore, a transmission configuration could open up possibilities for long distance quantum communication experiments with completely fibre-integrated components. 

\section*{methods}

%\subsection{Tapered Fibre Fabrication}
We fabricate tapered fibres using a heat-and-pull technique with a fixed hydrogen flame and two motorized stages. We use standard single mode fibre (Thorlabs 630HP), strip a 1 cm region of the outer jacket, and preheat the region for 120 s. We then pull the stages in opposite directions at a speed of 30 $\upmu$m s$^{-1}$. Laser transmission at 630 nm is monitored throughout the pulling process, and the stages are stopped manually when the transmission begins to fall. We confirmed with scanning-electron-micrscope and optical images that this diameter is on the order of 500 nm. 

Diamond micro-waveguides are detached and placed onto the tapered region using a tungsten micro-manipulator (Ted Pella). We use a rotation stage for the fibre to align it parallel to the micro-waveguide. The tungsten probe is held in a 3-axis piezo-controlled stage. 

%\subsection{Confocal Excitation and Autocorrelation Measurements}
A home-built confocal microscope is used to excite the NV centre. We use a 532 nm pump laser and focus it onto our sample using a Zeiss NA = 0.75 cover-glass corrected objective. We couple the signal from each fibre end to free-space in order to filter out the pump with two 530 nm long-pass filters. The resulting signal is coupled back into single-mode (SMF-28, Corning) fibre before being directed to two avalanche photo-diodes (Pelkin-Elmer). We perform cross-correlation measurements with a histogram of start-stop time intervals using a counting module with a bin width of 256 ns.

\begin{acknowledgments}
R.P. thanks Amir H. Safavi-Naeini for helpful discussion and Christopher Foy for useful comments on the manuscript. R.P. was supported in part by the MIT SuperUROP (Undergraduate Research Opportunities Program). T.S. was supported by the Alexander von Humboldt-Foundation. E.H.C. was supported by the NASA Office of the Chief Technologist's Space Technology Research Fellowship. Experiments were carried out in part with support from the Air Force Office of Scientific Research PECASE (supervised by G. Pomrenke). S.M. was supported by the AFOSR Quantum Memories MURI. Fabrication of the diamond micro-waveguides was carried out in part at the Center for Functional Nanomaterials, Brookhaven National Laboratory, which is supported by the US Department of Energy, Office of Basic Energy Sciences, under Contract No. DE-AC02-98CH10886.
% Put your acknowledgments here.
\end{acknowledgments}
\section*{Author Contributions}
R.P. , T.S., and N.W. performed the experiments. R.P. analysed the data and performed the numerical simulations. L.L. fabricated the diamond samples. S.M. and E.C. provided experimental assistance. R.P., T.S., and D.E. prepared the manuscript.

\bibliography{NWpaper_citations}

\begin{thebibliography}{10}
\expandafter\ifx\csname url\endcsname\relax
  \def\url#1{\texttt{#1}}\fi
\expandafter\ifx\csname urlprefix\endcsname\relax\def\urlprefix{URL }\fi
\providecommand{\bibinfo}[2]{#2}
\providecommand{\eprint}[2][]{\url{#2}}

\bibitem{Kimble2008}
\bibinfo{author}{Kimble, H.~J.}
\newblock \bibinfo{title}{{The quantum internet.}}
\newblock \emph{\bibinfo{journal}{Nature}} \textbf{\bibinfo{volume}{453}},
  \bibinfo{pages}{1023--30} (\bibinfo{year}{2008}).

\bibitem{Northup2014_Nature}
\bibinfo{author}{Northup, T.~E.} \& \bibinfo{author}{Blatt, R.}
\newblock \bibinfo{title}{{Quantum information transfer using photons}}.
\newblock \emph{\bibinfo{journal}{Nature Photonics}}
  \textbf{\bibinfo{volume}{8}}, \bibinfo{pages}{356--363}
  (\bibinfo{year}{2014}).

\bibitem{Childress2013_MRS}
\bibinfo{author}{Childress, L.} \& \bibinfo{author}{Hanson, R.}
\newblock \bibinfo{title}{{Diamond NV centers for quantum computing and quantum
  networks}}.
\newblock \emph{\bibinfo{journal}{MRS Bulletin}} \textbf{\bibinfo{volume}{38}},
  \bibinfo{pages}{134--138} (\bibinfo{year}{2013}).

\bibitem{Loncar2013_MRS}
\bibinfo{author}{Lon\v{c}ar, M.} \& \bibinfo{author}{Faraon, A.}
\newblock \bibinfo{title}{{Quantum photonic networks in diamond}}.
\newblock \emph{\bibinfo{journal}{MRS Bulletin}} \textbf{\bibinfo{volume}{38}},
  \bibinfo{pages}{144--148} (\bibinfo{year}{2013}).

\bibitem{sprague2014broadband}
\bibinfo{author}{Sprague, M.} \emph{et~al.}
\newblock \bibinfo{title}{Broadband single-photon-level memory in a hollow-core
  photonic crystal fibre}.
\newblock \emph{\bibinfo{journal}{Nature Photonics}}
  \textbf{\bibinfo{volume}{8}}, \bibinfo{pages}{287--291}
  (\bibinfo{year}{2014}).

\bibitem{thompson2013coupling}
\bibinfo{author}{Thompson, J.} \emph{et~al.}
\newblock \bibinfo{title}{Coupling a single trapped atom to a nanoscale optical
  cavity}.
\newblock \emph{\bibinfo{journal}{Science}} \textbf{\bibinfo{volume}{340}},
  \bibinfo{pages}{1202--1205} (\bibinfo{year}{2013}).

\bibitem{tiecke2014nanophotonic}
\bibinfo{author}{Tiecke, T.} \emph{et~al.}
\newblock \bibinfo{title}{Nanophotonic quantum phase switch with a single
  atom}.
\newblock \emph{\bibinfo{journal}{Nature}} \textbf{\bibinfo{volume}{508}},
  \bibinfo{pages}{241--244} (\bibinfo{year}{2014}).

\bibitem{TieckeFiber}
\bibinfo{author}{Tiecke, T.} \emph{et~al.}
\newblock \bibinfo{title}{Efficient fiber-optical interface for nanophotonic
  devices}.
\newblock \emph{\bibinfo{journal}{Optica}} \textbf{\bibinfo{volume}{2}},
  \bibinfo{pages}{70--75} (\bibinfo{year}{2015}).

\bibitem{Shea2013fiber}
\bibinfo{author}{O'Shea, D.}, \bibinfo{author}{Junge, C.},
  \bibinfo{author}{Volz, J.} \& \bibinfo{author}{Rauschenbeutel, A.}
\newblock \bibinfo{title}{Fiber-optical switch controlled by a single atom}.
\newblock \emph{\bibinfo{journal}{Physical review letters}}
  \textbf{\bibinfo{volume}{111}}, \bibinfo{pages}{193601}
  (\bibinfo{year}{2013}).

\bibitem{mitsch2014directional}
\bibinfo{author}{Mitsch, R.}, \bibinfo{author}{Sayrin, C.},
  \bibinfo{author}{Albrecht, B.}, \bibinfo{author}{Schneeweiss, P.} \&
  \bibinfo{author}{Rauschenbeutel, A.}
\newblock \bibinfo{title}{Directional nanophotonic atom--waveguide interface
  based on spin-orbit coupling of light}.
\newblock \emph{\bibinfo{journal}{arXiv preprint arXiv:1406.0896}}
  (\bibinfo{year}{2014}).

\bibitem{Maurer2012_science}
\bibinfo{author}{Maurer, P.~C.} \emph{et~al.}
\newblock \bibinfo{title}{{Room-temperature quantum bit memory exceeding one
  second.}}
\newblock \emph{\bibinfo{journal}{Science (New York, N.Y.)}}
  \textbf{\bibinfo{volume}{336}}, \bibinfo{pages}{1283--6}
  (\bibinfo{year}{2012}).

\bibitem{Bar-Gill2013_electronSpinCoherence}
\bibinfo{author}{Bar-Gill, N.}, \bibinfo{author}{Pham, L.~M.},
  \bibinfo{author}{Jarmola, A.}, \bibinfo{author}{Budker, D.} \&
  \bibinfo{author}{Walsworth, R.~L.}
\newblock \bibinfo{title}{Solid-state electronic spin coherence time
  approaching one second}.
\newblock \emph{\bibinfo{journal}{Nature communications}}
  \textbf{\bibinfo{volume}{4}}, \bibinfo{pages}{1743} (\bibinfo{year}{2013}).

\bibitem{Bernien2012_Nature_heraldedEntanglement}
\bibinfo{author}{Bernien, H.} \emph{et~al.}
\newblock \bibinfo{title}{{Heralded entanglement between solid-state qubits
  separated by 3 meters}}.
\newblock \emph{\bibinfo{journal}{Nature}} \textbf{\bibinfo{volume}{497}},
  \bibinfo{pages}{86--90} (\bibinfo{year}{2012}).
\newblock \eprint{1212.6136}.

\bibitem{pfaff2014unconditional}
\bibinfo{author}{Pfaff, W.} \emph{et~al.}
\newblock \bibinfo{title}{Unconditional quantum teleportation between distant
  solid-state qubits}.
\newblock \emph{\bibinfo{journal}{arXiv preprint arXiv:1404.4369}}
  (\bibinfo{year}{2014}).

\bibitem{Babinec2010}
\bibinfo{author}{Babinec, T.~M.} \emph{et~al.}
\newblock \bibinfo{title}{{A diamond nanowire single-photon source.}}
\newblock \emph{\bibinfo{journal}{Nature nanotechnology}}
  \textbf{\bibinfo{volume}{5}}, \bibinfo{pages}{195--9} (\bibinfo{year}{2010}).

\bibitem{schroder2011ultrabright}
\bibinfo{author}{Schr{\"o}der, T.}, \bibinfo{author}{G{\"a}deke, F.},
  \bibinfo{author}{Banholzer, M.~J.} \& \bibinfo{author}{Benson, O.}
\newblock \bibinfo{title}{Ultrabright and efficient single-photon generation
  based on nitrogen-vacancy centres in nanodiamonds on a solid immersion lens}.
\newblock \emph{\bibinfo{journal}{New Journal of Physics}}
  \textbf{\bibinfo{volume}{13}}, \bibinfo{pages}{055017}
  (\bibinfo{year}{2011}).

\bibitem{Hadden_etal}
\bibinfo{author}{Hadden, J.~P.} \emph{et~al.}
\newblock \bibinfo{title}{Strongly enhanced photon collection from diamond
  defect centers under microfabricated integrated solid immersion lenses}.
\newblock \emph{\bibinfo{journal}{Applied Physics Letters}}
  \textbf{\bibinfo{volume}{97}}, \bibinfo{pages}{--} (\bibinfo{year}{2010}).

\bibitem{Siyushev_etal}
\bibinfo{author}{Siyushev, P.} \emph{et~al.}
\newblock \bibinfo{title}{Monolithic diamond optics for single photon
  detection}.
\newblock \emph{\bibinfo{journal}{Applied Physics Letters}}
  \textbf{\bibinfo{volume}{97}}, \bibinfo{pages}{--} (\bibinfo{year}{2010}).

\bibitem{choy2013spontaneous}
\bibinfo{author}{Choy, J.~T.} \emph{et~al.}
\newblock \bibinfo{title}{Spontaneous emission and collection efficiency
  enhancement of single emitters in diamond via plasmonic cavities and
  gratings}.
\newblock \emph{\bibinfo{journal}{Applied Physics Letters}}
  \textbf{\bibinfo{volume}{103}}, \bibinfo{pages}{161101}
  (\bibinfo{year}{2013}).

\bibitem{li2014three}
\bibinfo{author}{Li, L.} \emph{et~al.}
\newblock \bibinfo{title}{Three megahertz photon collection rate from an nv
  center with millisecond spin coherence}.
\newblock \emph{\bibinfo{journal}{arXiv preprint arXiv:1409.3068}}
  (\bibinfo{year}{2014}).

\bibitem{Hausmann2012_nanoletters_NVringResonator}
\bibinfo{author}{Hausmann, B. J.~M.} \emph{et~al.}
\newblock \bibinfo{title}{{Integrated diamond networks for quantum
  nanophotonics.}}
\newblock \emph{\bibinfo{journal}{Nano letters}} \textbf{\bibinfo{volume}{12}},
  \bibinfo{pages}{1578--82} (\bibinfo{year}{2012}).

\bibitem{Chonan2014_natureSciReports}
\bibinfo{author}{Chonan, S.}, \bibinfo{author}{Kato, S.} \&
  \bibinfo{author}{Aoki, T.}
\newblock \bibinfo{title}{{Efficient single-mode photon-coupling device
  utilizing a nanofiber tip.}}
\newblock \emph{\bibinfo{journal}{Scientific reports}}
  \textbf{\bibinfo{volume}{4}}, \bibinfo{pages}{4785} (\bibinfo{year}{2014}).

\bibitem{almokhtar2014numerical}
\bibinfo{author}{Almokhtar, M.}, \bibinfo{author}{Fujiwara, M.},
  \bibinfo{author}{Takashima, H.} \& \bibinfo{author}{Takeuchi, S.}
\newblock \bibinfo{title}{Numerical simulations of nanodiamond nitrogen-vacancy
  centers coupled with tapered optical fibers as hybrid quantum nanophotonic
  devices}.
\newblock \emph{\bibinfo{journal}{Optics express}}
  \textbf{\bibinfo{volume}{22}}, \bibinfo{pages}{20045--20059}
  (\bibinfo{year}{2014}).

\bibitem{Yalla2012_PRL_QDonTaper}
\bibinfo{author}{Yalla, R.}, \bibinfo{author}{{Le Kien}, F.},
  \bibinfo{author}{Morinaga, M.} \& \bibinfo{author}{Hakuta, K.}
\newblock \bibinfo{title}{{Efficient Channeling of Fluorescence Photons from
  Single Quantum Dots into Guided Modes of Optical Nanofiber}}.
\newblock \emph{\bibinfo{journal}{Physical Review Letters}}
  \textbf{\bibinfo{volume}{109}}, \bibinfo{pages}{063602}
  (\bibinfo{year}{2012}).

\bibitem{Schroder2011_NanoLett_PCfiber}
\bibinfo{author}{Schr\"{o}der, T.}, \bibinfo{author}{Schell, A.~W.},
  \bibinfo{author}{Kewes, G.}, \bibinfo{author}{Aichele, T.} \&
  \bibinfo{author}{Benson, O.}
\newblock \bibinfo{title}{{Fiber-integrated diamond-based single photon
  source.}}
\newblock \emph{\bibinfo{journal}{Nano letters}} \textbf{\bibinfo{volume}{11}},
  \bibinfo{pages}{198--202} (\bibinfo{year}{2011}).

\bibitem{Schroder2012_NDonTaper_OptExp}
\bibinfo{author}{Schr\"{o}der, T.} \emph{et~al.}
\newblock \bibinfo{title}{{A nanodiamond-tapered fiber system with high
  single-mode coupling efficiency.}}
\newblock \emph{\bibinfo{journal}{Optics express}}
  \textbf{\bibinfo{volume}{20}}, \bibinfo{pages}{10490--7}
  (\bibinfo{year}{2012}).

\bibitem{Liebermeister2014}
\bibinfo{author}{Liebermeister, L.} \emph{et~al.}
\newblock \bibinfo{title}{{Tapered fiber coupling of single photons emitted by
  a deterministically positioned single nitrogen vacancy center}}.
\newblock \emph{\bibinfo{journal}{Applied Physics Letters}}
  \textbf{\bibinfo{volume}{104}}, \bibinfo{pages}{031101}
  (\bibinfo{year}{2014}).

\bibitem{Liu2013_APL}
\bibinfo{author}{Liu, X.} \emph{et~al.}
\newblock \bibinfo{title}{{Fiber-integrated diamond-based magnetometer}}.
\newblock \emph{\bibinfo{journal}{Applied Physics Letters}}
  \textbf{\bibinfo{volume}{103}}, \bibinfo{pages}{143105}
  (\bibinfo{year}{2013}).

\bibitem{fujiwara2011highly}
\bibinfo{author}{Fujiwara, M.}, \bibinfo{author}{Toubaru, K.},
  \bibinfo{author}{Noda, T.}, \bibinfo{author}{Zhao, H.-Q.} \&
  \bibinfo{author}{Takeuchi, S.}
\newblock \bibinfo{title}{Highly efficient coupling of photons from
  nanoemitters into single-mode optical fibers}.
\newblock \emph{\bibinfo{journal}{Nano letters}} \textbf{\bibinfo{volume}{11}},
  \bibinfo{pages}{4362--4365} (\bibinfo{year}{2011}).

\bibitem{Groblacher2013_adiabaticCoupling_APL}
\bibinfo{author}{Groblacher, S.}, \bibinfo{author}{Hill, J.~T.},
  \bibinfo{author}{Safavi-Naeini, A.~H.}, \bibinfo{author}{Chan, J.} \&
  \bibinfo{author}{Painter, O.}
\newblock \bibinfo{title}{{Highly efficient coupling from an optical fiber to a
  nanoscale silicon optomechanical cavity}}.
\newblock \emph{\bibinfo{journal}{Applied Physics Letters}}
  \textbf{\bibinfo{volume}{103}}, \bibinfo{pages}{181104}
  (\bibinfo{year}{2013}).

\bibitem{fiberTapering}
\bibinfo{author}{{Hauer}, B.~D.} \emph{et~al.}
\newblock \bibinfo{title}{{On-Chip Cavity Optomechanical Coupling}}.
\newblock \emph{\bibinfo{journal}{ArXiv e-prints}}  (\bibinfo{year}{2014}).
\newblock \eprint{1401.5482}.

\bibitem{Vetsch2010_laserTrappedNanofiber_PRL}
\bibinfo{author}{Vetsch, E.} \emph{et~al.}
\newblock \bibinfo{title}{{Optical Interface Created by Laser-Cooled Atoms
  Trapped in the Evanescent Field Surrounding an Optical Nanofiber}}.
\newblock \emph{\bibinfo{journal}{Physical Review Letters}}
  \textbf{\bibinfo{volume}{104}}, \bibinfo{pages}{203603}
  (\bibinfo{year}{2010}).

\bibitem{liSciReportsDiamond2015}
\bibinfo{author}{Li, L.} \emph{et~al.}
\newblock \bibinfo{title}{Nanofabrication on unconventional substrates using
  transferred hard masks}.
\newblock \emph{\bibinfo{journal}{Sci. Rep.}} \textbf{\bibinfo{volume}{5}}
  (\bibinfo{year}{2015}).

\bibitem{dolan2014complete_polarizationRadiallyPolarizedBeams_NV}
\bibinfo{author}{Dolan, P.~R.}, \bibinfo{author}{Li, X.},
  \bibinfo{author}{Storteboom, J.} \& \bibinfo{author}{Gu, M.}
\newblock \bibinfo{title}{Complete determination of the orientation of nv
  centers with radially polarized beams}.
\newblock \emph{\bibinfo{journal}{Optics express}}
  \textbf{\bibinfo{volume}{22}}, \bibinfo{pages}{4379--4387}
  (\bibinfo{year}{2014}).

\bibitem{PhysRevB.76.165205_PolarizationSelectiveExcitation_AlegreSantori}
\bibinfo{author}{Alegre, T. P.~M.}, \bibinfo{author}{Santori, C.},
  \bibinfo{author}{Medeiros-Ribeiro, G.} \& \bibinfo{author}{Beausoleil, R.~G.}
\newblock \bibinfo{title}{Polarization-selective excitation of nitrogen vacancy
  centers in diamond}.
\newblock \emph{\bibinfo{journal}{Phys. Rev. B}} \textbf{\bibinfo{volume}{76}},
  \bibinfo{pages}{165205} (\bibinfo{year}{2007}).

\bibitem{saraNWwaveguide}
\bibinfo{author}{{Mouradian}, S.~L.} \emph{et~al.}
\newblock \bibinfo{title}{{The Scalable Integration of long-lived quantum
  memories into a photonic circuit}}.
\newblock \emph{\bibinfo{journal}{ArXiv e-prints}}  (\bibinfo{year}{2014}).
\newblock \eprint{1409.7965}.

\bibitem{li2014coherent}
\bibinfo{author}{Li, L.} \emph{et~al.}
\newblock \bibinfo{title}{Coherent spin control of a nanocavity-enhanced qubit
  in diamond}.
\newblock \emph{\bibinfo{journal}{Nature Communications}}
  (\bibinfo{year}{2015}).

\bibitem{wgcodeMurphy}
\bibinfo{author}{Fallahkhair, A.~B.}, \bibinfo{author}{Li, K.~S.} \&
  \bibinfo{author}{Murphy, T.~E.}
\newblock \bibinfo{title}{Vector finite difference modesolver for anisotropic
  dielectric waveguides}.
\newblock \emph{\bibinfo{journal}{J. Lightw. Technol.}}
  \textbf{\bibinfo{volume}{26}}, \bibinfo{pages}{1423--1431}
  (\bibinfo{year}{2008}).

\end{thebibliography}

\section*{Supplementary Information}
\subsection{Experimental Setup and Sample Preparation}

\begin{figure*}
 \centerline{\includegraphics[scale=.45]{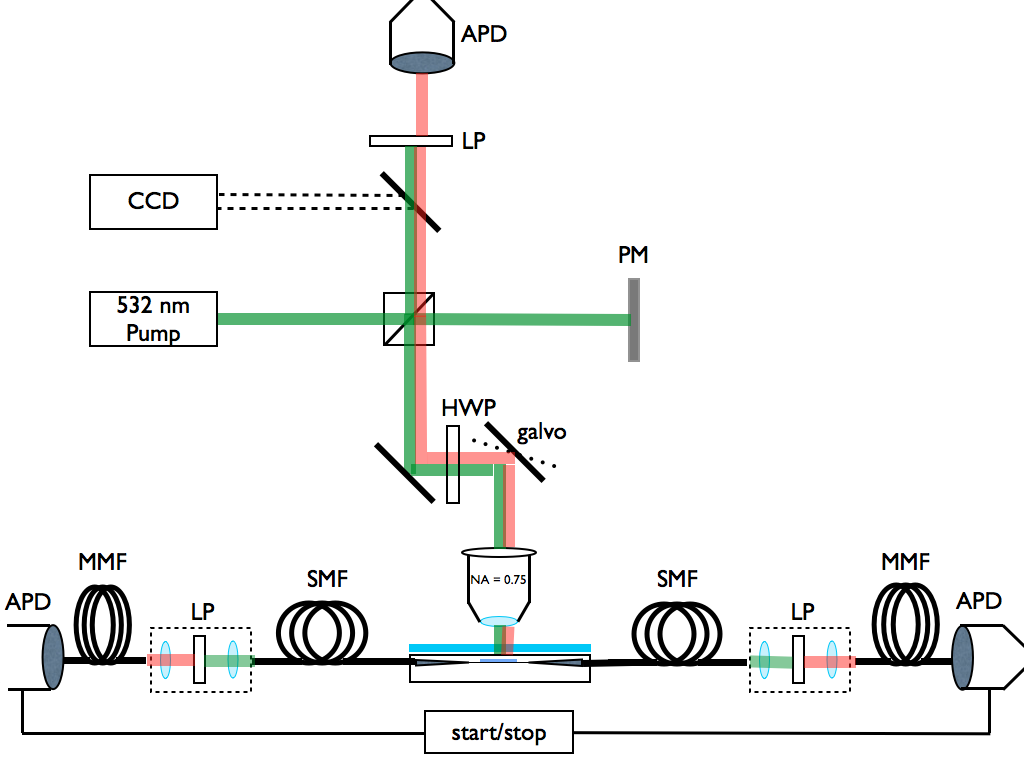}}%
 \caption{\textbf{Experimental Setup} Illustration of confocal setup used to perform optical excitation and collection measurements. The abbreviations are APD: Avalanche photo diode, LP: Long pass filter, CCD: white light camera, PM: power meter, HWP: half wave plate, galvo: galvanometer, SMF: single mode fibre, MMF: multi mode fibre.}%
 \label{fig:illustration}
 \end{figure*}

	The detailed experimental setup is shown in Fig. \ref{fig:illustration}. In this confocal setup, a 532 nm laser is scanned using two galvanometers over the sample. The resulting fluorescence is filtered with a 550 nm long pass filter to eliminate the green pump, and detected using avalanche photo diodes (APDs). Note that light going to any APD in the figure can equivalently be passed to a spectrometer. Multimode fibre is used after the long pass filtering stages to minimize losses. 
	
	Our fibre taper fabrication setup uses two motorized stages pulling in opposite directions at a speed of 30 $\upmu$m/s. We use a hydrogen gas flame to heat an approximately 1cm segment of a standard single mode fibre. The region to be heated is stripped of its plastic jacket before heating. The pulling distance is calibrated by monitoring the transmission of a 635 nm laser. The process is stopped manually when the transmission begins to fall rapidly, allowing for $<$ 500 nm taper diameter with transmission $>$ 90\%. To secure the fibre, we use UV adhesive glue and a curing lamp station. The fibre sits in a 500 $\upmu$m deep groove in a metal holder. A glass cover-slip is placed directly above, to protect the fibre and allow optical excitation. 
	
	For assembly of the fibre-taper micro-waveguide system, we use a tungsten probe (Tedpella 0.6 $\upmu$m tip radius) to pick up and place individual diamond micro-waveguides onto the tapered region. We place the fibre mount on a rotating stage and image the fibre using a high working distance homemade optical microscope. We move the tungsten probe, with a three axis piezoelectric micrometer system, into the same field of view, and detach a single micro-waveguide onto the taper surface. This is achieved by bringing the tip in contact with the taper, after aligning the taper such that it is parallel with the diamond micro-waveguide. 
	
\subsection{Spin Coherence in Diamond Micro-waveguide Structures}
\begin{figure*}
 \centerline{\includegraphics[scale=.2]{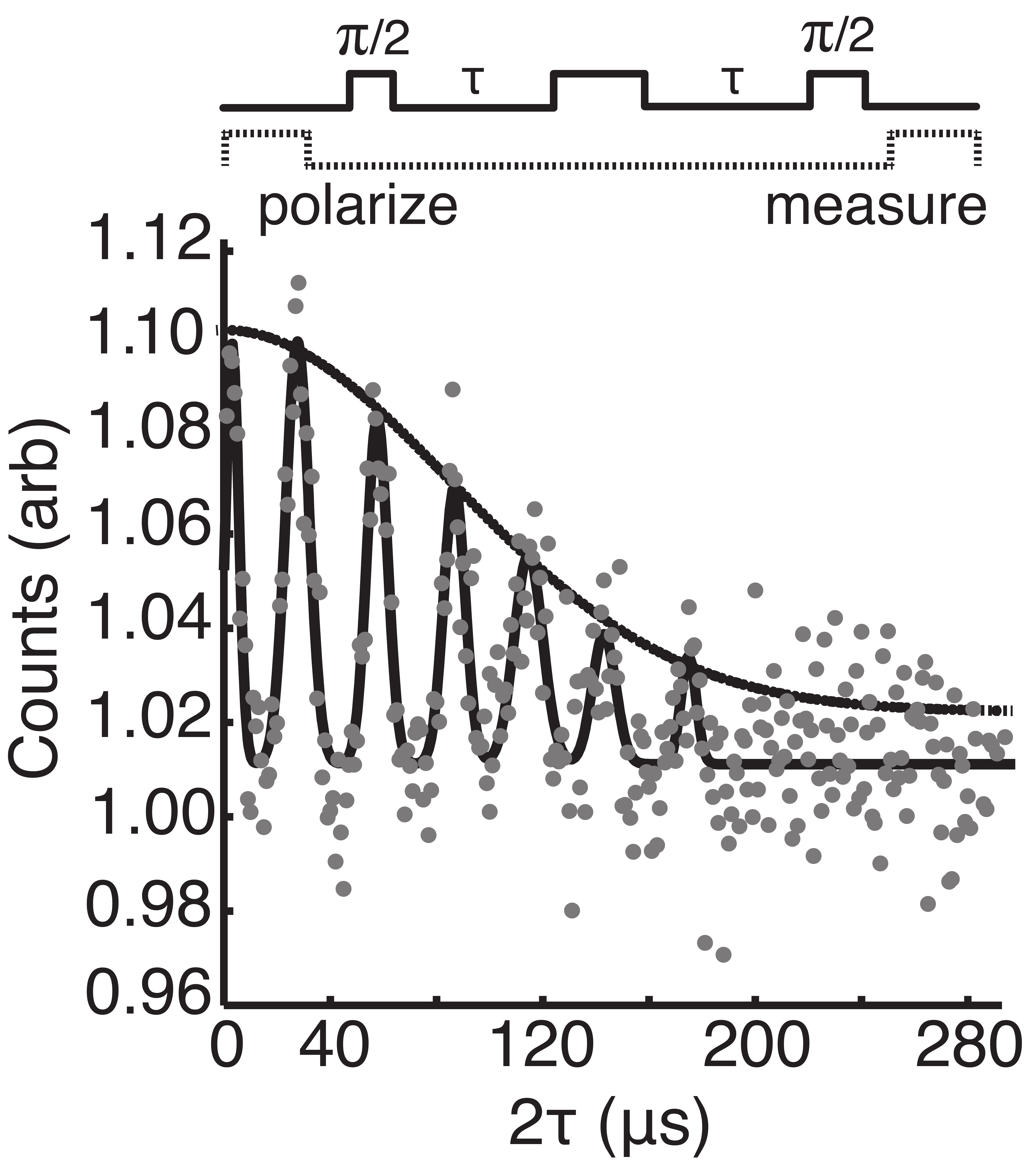}}%
 \caption{\textbf{Hahn Echo of NV electron spin in Diamond Micro-waveguide} Hahn-echo showing $T_{2} > 120 \upmu$s \cite{saraNWwaveguide} }%
 \label{fig:echo}
 \end{figure*}

To show the suitability of our diamond micro-waveguides as quantum nodes, we performed spin measurements on equivalent micro-waveguides containing NVs. Measurements of the electron spin coherence time using a Hahn-Echo pulse sequence in related diamond micro-waveguide structures show $T_{2} > 120 \upmu s$ (Fig. \ref{fig:echo}) \cite{saraNWwaveguide}. 

\subsection{NV Orientation}

\begin{figure*}
 \centerline{\includegraphics[scale=.45]{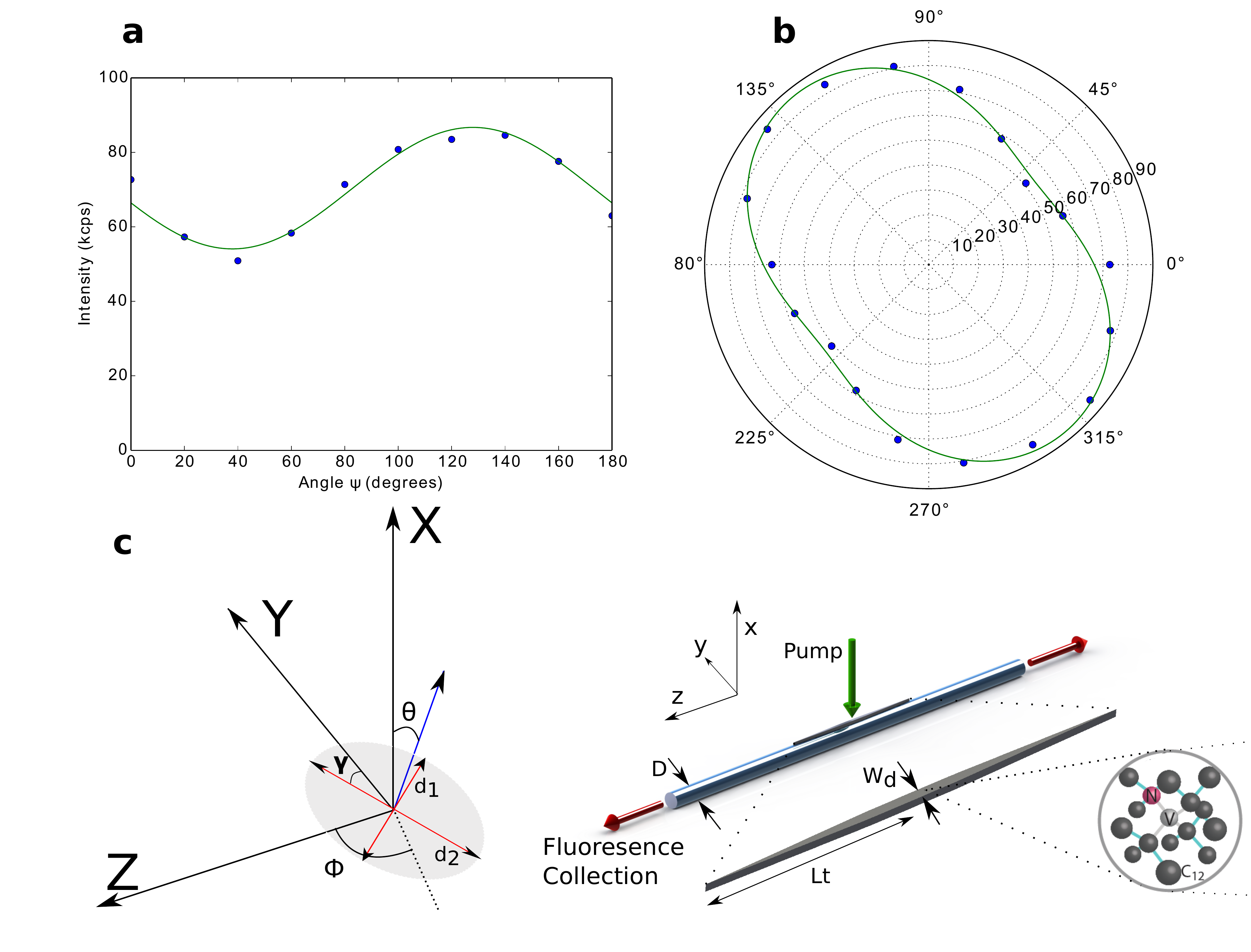}}%
 \caption{\textbf{Excitation Polarization Sweep} \textbf{(a)} Polarization dependent fluorescence data for a single NV in a diamond micro-waveguide. \textbf{(b)} Same data as in (a) but in polar form. \textbf{(c)} Coordinate inset showing NV symmetry axis (blue) and NV radiation dipoles $d_{1}$ and $d_{1}$. Note that, as in the main text, our Cartesian system is oriented such that Z is along the fibre, and the X axis points to the objective.}%
 \label{fig:polarization}
 \end{figure*}

	To determine the orientation of an NV centre relative to the micro-waveguide, we rotate the polarization angle $\psi$ of a linearly polarized pump beam, and record the resulting fibre-coupled fluorescence. Such a method is analogous to those in \cite{Liebermeister2014}\cite{dolan2014complete_polarizationRadiallyPolarizedBeams_NV}\cite{PhysRevB.76.165205_PolarizationSelectiveExcitation_AlegreSantori}. The results are given in Fig. \ref{fig:polarization}, which shows the raw polarization dependent intensity data, as well as a fit to the function \cite{Liebermeister2014}: 
	\begin{equation}
	I = I_o (\sin^{2}(\psi - \phi) + \cos^{2}(\psi - \phi)\cos^{2}(\theta)) 
	\end{equation}
	where the angles $\phi$ and $\theta$ denote the direction of the NV symmetry axis in spherical polar coordinates. The fit yields $\theta = 37.8 \pm 1.7 \degree$ and $\phi = 38.0 \pm 2.5 \degree$. The two NV radiation dipoles are perpendicular to each other, and are oriented in a plane perpendicular to the NV symmetry axis. The orientation angle $\gamma$ of the NV dipoles within this plane cannot be determined from our measurement. 
	
\subsection{Numerical Simulations}
We use a freely available MATLAB code developed in \cite{wgcodeMurphy} to calculate the band diagrams in the main text (main text Fig. 2a). The fundamental anti-symmetric mode is plotted in the main text. The corresponding mode fields (main text Fig. 2b) are plotted using COMSOL, for the corresponding effective mode indices returned by the MATLAB code. 

We use the Lumerical FDTD solution package to calculate the expected collection efficiency from a diamond micro-waveguide coupled to an optical fibre. The micro-waveguide is specified as three geometry blocks consisting of two triangular tapered ends, and one central region of constant length. The diamond index is specified as 2.42. To sweep the length, we change the length of the two tapered ends. The tapered fibre segment is modeled as a dielectric cylinder with a refractive index of 1.52. 

We monitor the flux through eight flux planes. Six flux planes define a closed box surrounding the entire structure. They are used to calculate the total power emitted by a dipole source. Two square flux planes of 500 nm length are placed at the fibre ends. The collection efficiency is specified by the net optical power passing through the flux planes on the fibre with an added correction to take into account the power transported by the evanescent fields (in the wave guiding direction). This power fraction is determined from a finite-element method (COMSOL) and we determine that 91 percent of the optical power travels within the fibre taper with diameter of 500 nm. Finally, the collection efficiency is defined by the ratio of the guided flux to the total flux, all divided by the factor 0.91. 

We use a dipole source that generates a gaussian pulse for our optical excitation in the FDTD simulation. For simplicity, we simulate an idealized system, and place the source at the centre of the micro-waveguide. Each simulation generates data at 50 frequency points between 538.4 nm and 738.5 nm. For varying diamond waveguide lengths, we calculate the collection efficiency centered at 637 nm over a range from 563 nm to 733 nm. 

\subsection{Determination of Collection Efficiency from Realistically Oriented NV dipoles}

Here, we proceed to use our knowledge of the NV orientation and the results of FDTD simulations to estimate two numbers. The first is the theoretical value for the collection efficiency $\eta_{c}$, which requires a projection of the dipole orientation onto the three basis polarizations for which FDTD data is taken. The second is the representative fraction of photons collected into the acceptance angle of the microscope, $\kappa_{NA}$ which is necessary to calculate $\Gamma_{free}$, the total number of photons emitted into free space. $\Gamma_{free}$ is used in the main text to determine an experimental value of $\eta_{c}$ and through the relation $\eta_{c} = \beta / \eta_{QE}$ to estimate the quantum efficiency. Our methods are similar to an approach described in \cite{Liebermeister2014}. 

To determine the expected collection efficiency consistent with the experimental parameters, we must take into account the NV dipole orientations. We obtain the collection efficiencies from FDTD simulations of a single dipole along X,Y, and Z directions. We then take the projection of the NV dipoles onto these axes, and determine the collection efficiency for a micro-waveguide of length 12 um.

To obtain the dipole projections onto the X,Y, and Z axes we note that the NV dipoles are oriented according to the angles $\theta$, $\phi$, and $\gamma$ (see Fig. \ref{fig:polarization}). $\gamma$ is undetermined from our measurements, so in the following discussion we sweep $\gamma$ over $360\degree$. For the other angles we use $\phi = 38 \degree$ and $\theta = 37.8 \degree$. The projections are obtained by applying:

\begin{equation}
\mathbf{d_{k final}} = R_{X}(\phi) R_{Z}(\theta) R_{X} (\gamma) \mathbf{d_{k initial}}
\end{equation}
where $k = 1,2$. $\mathbf{d_{1initial}}$ points along the Z axis and $\mathbf{d_{2initial}}$ along the Y axis. $R_{X}$ generates a rotation about the X axis. $R_{Z}$ generates a rotation about the Z axis. The components of $\mathbf{d_{kfinal}}$ ($d_{kX}$,$d_{kY}$,$d_{kZ}$) contain the desired projection. 
 
We denote the simulated collection efficiencies for a dipole oriented in the X,Y, and Z directions as $\eta_{X}, \eta_{Y}$, and $\eta_{Z}$ respectively. Then the collection efficiency from each NV dipole is computed as:
\begin{equation}
\eta_{ck} = \sqrt{(\eta_{X} d_{kX})^{2} +  (\eta_{Y} d_{kY})^{2} + (\eta_{Z} d_{kZ})^{2}}
\end{equation}
for k = 1,2 and $\eta_{X},\eta_{Y},\eta_{Z} = 0.71,0.63,0.04$. The total efficiency is the average of the two dipole contributions: 
\begin{equation}
\eta_{c} = \frac{\eta_{c1} + \eta_{c2}}{2}
\end{equation}

By variation of $\gamma$, we obtain the following range of values for the theoretical collection efficiency: $0.49 < \eta_{c} < 0.52$.

From FDTD calculations, we determined the $\kappa_{NA}$ for the X,Y,Z directions by calculating the fraction of the free space power radiated into an objective of NA  = 0.75. This corresponds to an acceptance cone of half angle $48.6 \degree$. The calculation is performed using the Lumerical FDTD \textbf{farfield3dintegrate} routine and correcting for the guided power (since we are interested in the fraction of the free space power). 

In exactly the same way as before, we calculate $\kappa_{k}$ (dropping the NA subscript for clarity) using all three dipole components: 

\begin{equation}
\kappa_{k} = \sqrt{(\kappa_{X} d_{kX})^{2} +  (\kappa_{Y} d_{kY})^{2} + (\kappa_{Z} d_{kZ})^{2}}
\end{equation}

for $k = 1,2$ and $\kappa_{X}, \kappa_{Y}, \kappa_{Z} = 0.09, 0.22, 0.21$. The total efficiency is: 
\begin{equation}
\kappa = \frac{\kappa_{1} + \kappa_{2}}{2}
\end{equation}

By variation of $\gamma$, we determine: $0.197 < \kappa_{NA} < 0.198 $
\end{document}